\documentclass[conference,10pt]{IEEEtran}
\usepackage{epsfig,epsf,rotating,setspace,latexsym,amsmath,amssymb,amsfonts,bm,theorem,subfigure,epstopdf,cite,authblk,bbm,nonfloat,comment}
\usepackage{algorithm}
\usepackage[noend]{algpseudocode}
\usepackage{color}
\usepackage{mathtools}
\usepackage{soul}

\algrenewcommand\algorithmicforall{\textbf{foreach}}
\algrenewcommand\algorithmicindent{.8em}

\newtheorem{theorem}{Theorem}
\newtheorem{lemma}{Lemma}
\newtheorem{definition}{Definition}
\newtheorem{corollary}{Corollary}

\newenvironment{Proof}[1]{\medskip\par\noindent{\bf Proof:\,}\,#1}{{\mbox{\,$\blacksquare$}\par}}
\newenvironment{ProofLemma1}[1]{\medskip\par\noindent{\bf Proof of Lemma~\ref{lemma: 1}:\,}\,#1}{{\mbox{\,$\blacksquare$}\par}}
\newenvironment{ProofLemma2}[1]{\medskip\par\noindent{\bf Proof of Lemma~\ref{lemma: 2}:\,}\,#1}{{\mbox{\,$\blacksquare$}\par}}

\IEEEoverridecommandlockouts
\allowdisplaybreaks
\begin{document}

\title{ Information Freshness in Dynamic Gossip Networks} 

\author{Arunabh Srivastava \qquad Thomas Jacob~Maranzatto \qquad Sennur Ulukus\\
        \normalsize Department of Electrical and Computer Engineering\\
        \normalsize University of Maryland, College Park, MD 20742\\
        \normalsize  \emph{arunabh@umd.edu} \qquad \emph{tmaran@umd.edu} \qquad \emph{ulukus@umd.edu}}

\maketitle

\begin{abstract}
    We consider a source that shares updates with a network of $n$ gossiping nodes. The network's topology switches between two arbitrary topologies, with switching governed by a two-state continuous time Markov chain (CTMC) process. Information freshness is well-understood for static networks. This work evaluates the impact of time-varying connections on information freshness. In order to quantify the freshness of information, we use the version age of information metric. If the two networks have static long-term average version ages of $f_1(n)$ and $f_2(n)$ with $f_1(n) \ll f_2(n)$, then the version age of the varying-topologies network is related to $f_1(n)$, $f_2(n)$, and the transition rates in the CTMC. If the transition rates in the CTMC are faster than $f_1(n)$, the average version age of the varying-topologies network is $f_1(n)$. Further, we observe that the behavior of a vanishingly small fraction of nodes can severely impact the long-term average version age of a network in a negative way. This motivates the definition of a typical set of nodes in the network. We evaluate the impact of fast and slow CTMC transition rates on the typical set of nodes. 
\end{abstract}

\section{Introduction}\label{sec: introduction}
Wireless connectivity is becoming increasingly integral to day-to-day activities involving both humans and machines. Following this change, networks with dynamic and adaptive configurations have gained significant interest. The rapid evolution of wireless communication technologies, with the advent of the 5G communication standards and the emerging 6G communication standards, has introduced new paradigms such as ultra-reliable low latency communication (URLLC) and massive machine type communication (mMTC) \cite{pokhrel2020}. These advancements have contributed to the growth of internet of things (IoT) networks, including a range of applications such as environmental sensor networks, industrial automation and drone formations. These networks operate under time-sensitive constraints, and require efficient and adaptive information dissemination strategies for the network.

A key challenge in such time-sensitive networks is ensuring timely and decentralized information exchange between nodes. Centralized architectures are unable to overcome this challenge due to scalability limitations, communication delays and a lack of robustness to points of failure. For example, in a swarm drone formation where drones perform coordinated maneuvers, drones need to share their exact positions with neighbors continuously. Performing this task with a centralized controller may lead to delays and bottlenecks, as the network scales or experiences interference. This issue can be addressed using gossiping protocols\cite{chettri2019comprehensive, swamy2020empirical}, which offer a decentralized solution. In a gossiping protocol, nodes communicate locally to propagate information across the network. These protocols enable scalable, efficient and fault-tolerant information dissemination in dynamic environments.

In addition, real-time decision making in dynamic and time-sensitive networks hinges on the freshness of shared information. Stale or outdated information can lead to inefficiencies and errors, especially in time-critical applications such as autonomous vehicle coordination or process control in smart factories. In order to use fresh information for quantitative decision-making, the age of information (AoI) metric has been widely adopted\cite{kaul2012real, sun2019age, yatesJSACsurvey}. AoI  measures the time elapsed since the last successfully received update, and several works have utilized the AoI metric to measure network performance \cite{yates2018age, banerjee2023re}. This has led to the development of several performance-based freshness metrics, including the age of incorrect information \cite{maatouk20AOII}, the age of synchronization \cite{zhong18AoSync}, binary freshness metric \cite{cho3BinaryFreshness}, and the version age of information \cite{yates21gossip, Abolhassani21version, melih2020infocom}.

In this work, we use the version age of information metric to quantify information freshness. A difference between AoI and version age is that version age is discrete. This metric is particularly relevant for gossiping networks where updates are timestamped, thus a discrete measure of freshness is needed. The version age of information of a node in a gossiping network is defined as the difference between the source version and the node version. This model was first analyzed using the spatial mean field regime method in \cite{chaintreau2009age}, and later using the stochastic hybrid system (SHS) framework in \cite{yates21gossip}. Following works \cite{buyukates22ClusterGossip, kaswan22jamming, mitra_Infocom23, srivastava2024mobilityagebasedgossipnetworks, maranzatto24} have focused on using the SHS framework and the resulting recursive equations to analyze gossiping networks with diverse properties, including structured topologies, mobility, and adversarial settings. A comprehensive review of these works is provided in \cite{kaswan2023versionagesurvey}. Additionally, \cite{maranzatto2024agegossipconnectiveproperties} characterized the distribution of the version age of information using first passage percolation.

Building upon these prior works, this paper investigates gossiping networks with time-varying topologies where the underlying network architecture evolves dynamically. We aim to characterize the effect of network dynamics on the long-term average version age of the entire network. Specifically, we consider networks with a finite set of fixed topologies, where transitions between topologies follow a continuous time Markov chain (CTMC) process. A restricted model was explored in \cite{srivastava2024varyingtopologies}, where one of the topologies is forced to be the fully-connected topology. There, the version age of each node in the gossiping network remains logarithmic if the CTMC rates are constant; this matches the performance of a static fully-connected network. Our main contributions are twofold. First, we allow switching between arbitrary topologies. Second, we analyze a broad spectrum of CTMC transition rates. In addition to this, we establish fundamental properties of single topology networks in Section~\ref{sec: fast switching}. Our findings in Section~\ref{sec: typical sets} reveal that a small subset of nodes can disproportionately affect the version age of the entire network. As a first, we introduce the concept of a typical set of nodes in gossiping networks in this work. We prove results for varying-topology networks which hold for all but a small subset of nodes for a large spectrum of CTMC rates between $0$ and $\infty$.

\begin{figure}
    \centering
    \includegraphics[width=0.75\linewidth]{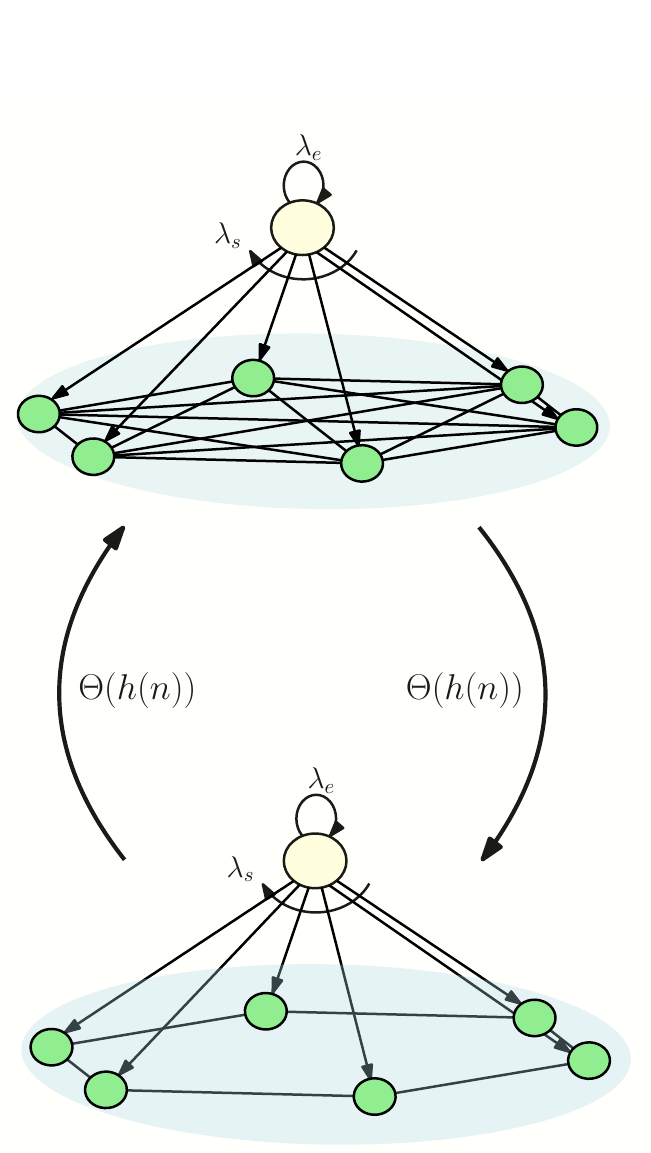}
    \caption{An example of a dynamic network we study. Here, $n=6$, topology 1 is the fully-connected topology, and topology 2 is the ring topology. We have omitted the edge-rates, but these implicitly change as the network switches between topologies 1 and 2.}
    \label{fig:dynamics_pic}
\end{figure}

\section{System Model and the Version Age Metric}\label{sec: system model}
We consider a source node $n_0$ sending updates to a set of nodes $\mathcal{N} = [1,\ldots, n]$. The source updates itself via a Poisson process with rate $\lambda_e$. The source also sends updates to every $j \in \mathcal{N}$ via independent Poisson processes each with rate $\frac{\lambda_s}{n}$. At any time $t \geq 0$, nodes in $\mathcal{N}$ can send updates to their neighbors. However, the set of neighbors for a node may be different at different times. We model the dynamics of the edge set by considering a finite set of networks $\mathcal{K} = \{1,2,\ldots,K\}$ where $K < \infty$. We refer to $\mathcal{K}$ as the \textit{set of topologies} for the system. We view $\mathcal{K}$ as the state space for a continuous time Markov chain (CTMC), with transition rates $q_{kl}$ between states $k\in \mathcal{K},l \in \mathcal{K}$. We also assume that the holding time in each state of the Markov chain, defined as the random variable associated with the time it takes to leave the state, is of the same order as a function $h(n)$ of $n$. This means that the CTMC stays in a state for a time that is exponentially distributed with mean of the same order as $h(n)$.

When the network is in topology $k$, node $i$ sends updates to node $j$ via another independent Poisson process with rate $\lambda_{i,j}^{(k)}$. We assume all nodes communicate with the same constant rate. That is, setting $N^{(k)}(i)$ to be the neighbors of $i$ in topology $k$, we assume $\sum_{j \in N^{(k)}(i)}\lambda_{i,j}^{(k)} = \lambda$, for all $i\in \mathcal{N}$ and $k \in \mathcal{K}$. We assume $\mathcal{K}$ is constant as $n \to \infty$, but that the transition rates and probabilities can depend on $n$. We do not assume that the rates are equal among all neighbors, that is, $\lambda_i(j) = \frac{\lambda}{\operatorname{outdeg}(i)}$ may not hold in general. We also note that the process $\lambda_{i,j}$ can be modeled as a doubly-stochastic/Cox process, though we do not use this characterization for the analysis. For the rest of the paper we set $\lambda_s = \lambda$.

Thus far, we have only discussed how updates are sent, not the messages contained in each update. In our model, the source and every node in the network contain internal counters which represent versions of some update; when a node $i \in \mathcal{N} \cup {n_0}$ communicates to a neighbor $j$ (because $i$’s Poisson process updated), $i$ sends its current version. The version for $n_0$ increments if and only if the process for $n_0$ updates. That is, the counter in $n_0$ is exactly a Poisson counting process with intensity $\lambda_e$. A node $j \in \mathcal{N}$ updates its version at time $t$ if and only if some other node with a larger version sends a message to $j$ at time $t$. Letting $N_i(t)$ be the counting process associated with vertex $i \in n_0 \cup \mathcal{N}$, the version age of node $i$ at time $t$ is $X_i(t) = N_0(t) - N_i(t)$. If $G$ is a static network, we write $X_i^{G}(t)$ to mean the version age of node $i$ at time $t$ in network $G$. We define the long-term average version age of a varying-topologies network as $v=\lim_{t \rightarrow\infty}\frac{1}{n} \sum\limits_{i \in \mathcal{N}}X_i(t)$.

Throughout we assume $\lambda_e$, $\lambda$ are constants and let $n \rightarrow \infty$. We say that an event $\mathcal{A}$ holds with high probability (w.h.p.) if $\mathbb{P}[\mathcal{A}] \to 1$ as $n \to \infty$. We use the standard big-O definitions. We say that $f(n) = O(g(n))$ if $\limsup_{n \to \infty} \frac{f(n)}{g(n)} < \infty$, $f(n) = \Omega(g(n))$ if $\limsup_{n \to \infty} \frac{g(n)}{f(n)} < \infty$, $f(n) = \Theta(g(n))$ if $c_1 \leq \lim_{n \to \infty}\frac{f(n)}{g(n)} \leq c_2$ for some constants $c_1 \leq c_2$, $f(n) = o(g(n))$ if $\lim_{n \to \infty} \frac{f(n)}{g(n)} = 0$ and $f(n) = \omega(g(n))$ if $g(n) = o(f(n))$. We denote sets using calligraphic notation.

Our analysis is stated in the form of infinite network sequences of the form $\{G_n\}_{n=1}^\infty$, where network $G_k$ has $k$ nodes. Other works in gossiping have implicitly used graph sequences in their proofs, but since the networks considered had nice properties, such as vertex transitivity (e.g., the generalized rings networks in~\cite{srivastava2023generalizedrings}) or easy parametrization (e.g., the Erdos-Renyi graph in~\cite{maranzatto24}), the presentation using graph sequences could be replaced with more straightforward methods. Our primary objective is to understand what happens when two large networks which have version ages scaling as $\Theta(f_1(n))$ and $\Theta(f_2(n))$ interact via a CTMC with holding times $h(n)$ below $f_1(n)$, above $f_2(n)$, and intermediate to $f_1(n)$ and $f_2(n)$. Thus, we choose $K=2$ in the rest of the paper. In this setting we state our results in the form of graph sequences, though we are mostly considering well-behaved sequences chosen from a natural family such as the graphs described in the papers above.

In this paper, we consider a running example where the topologies switch between a fully-connected and a ring network. We use this example to illustrate our results in a straightforward manner.

\section{Fast Switching}\label{sec: fast switching}
Suppose we have two distinct network topologies, topology $1$ and topology $2$, represented using $G_1$ and $G_2$, respectively. Informally, suppose the long-term average version ages of $G_1$ and $G_2$ are $\Theta(f_1(n))$ and $\Theta(f_2(n))$, respectively. It follows from the bounds in \cite{yates21gossip} that we can take the rates $\Theta(\log n) \leq f_1(n),f_2(n) \leq \Theta(n)$. Furthermore assume that $f_1(n) = o(f_2(n))$, and suppose the CTMC rates satisfy $q_{12}, q_{21} = \Theta(1/h(n))$ and holding times are $\Theta(h(n))$. An example of this setup with holding times is illustrated in Fig.~\ref{fig:dynamics_pic}. In this section, we characterize the long-term average version age of the network when $h(n)$ varies between $0$ and $\Theta(f_1(n))$. For example, if the transition rates are constants then it is very unlikely an update will reach every node without a CTMC update occurring. The behavior of the fast switching process is studied in Theorem~\ref{thm: 1}. 

We need the following lemmas, with proofs deferred to the Appendix. Lemma~\ref{lemma: 1} shows that any gossiping network is able to maintain the version age of all nodes in the network at or below its long-term average version age. Lemma~\ref{lemma: 2} shows that we spend the same amount of time in both states in order as a function of $n$. Lemma~\ref{lemma: 3} shows that the number of times source updates itself, i.e., generates a new update, is of the same order as the time that has elapsed. The proof for Lemma~\ref{lemma: 2} follows along similar lines as the proof of \cite[Lemma~2]{srivastava2024varyingtopologies} and Lemma~\ref{lemma: 3} is a simple application of Chebyshev's inequality.

\begin{lemma}\label{lemma: 1}
    Let $\{G_n\}_{n=1}^\infty$ be a sequence of networks with $|\mathcal{N}(G_k)| = k$.  Suppose $\lim\limits_{t \to \infty} \mathbb{E} [X_1^{G_n}(t)] = \Theta(f(n))$. Then, for large enough $n$ and $t$, with high probability, $X_1^{G_n}(t) = O(f(n))$.
\end{lemma}

\begin{lemma}\label{lemma: 2}
    If the CTMC has spent $\Theta(h(n))$ time in state $k \in \{1,2\}$ at time $T$, then with high probability, $T = \Theta(h(n))$.
\end{lemma}

\begin{lemma}\label{lemma: 3}
    Suppose $\frac{\lambda_e}{\lambda}\log{n} \leq f(n) \leq \frac{\lambda_e}{\lambda}n$. If $T = \Theta(f(n))$, then with high probability, $N_0([0,T]) = \Theta(f(n))$.
\end{lemma}

We are now ready to prove the main result of this section.

\begin{theorem}\label{thm: 1}
    Let $\{G_{1,n}\}_{n=1}^\infty$ and $\{G_{2,n}\}_{n=1}^\infty$ be two sequences of networks with $|\mathcal{N}(G_{1,k})| = |\mathcal{N}(G_{2,k})| = k$. Suppose $G_{1,n}$ has a long-term average version age $\Theta(f_1(n))$ and $G_{2,n}$ has a long-term average version age $\Theta(f_2(n))$. Further, let $f_1(n) = o(f_2(n))$. Let the transition rates have the same order, i.e., $q_{12}, q_{21} = \Theta(1/h(n))$. If $h(n) = O(f_1(n))$, then the long-term average version age for the time-varying system is $\Theta(f_1(n))$.
\end{theorem}

The proof of Theorem~\ref{thm: 1} is a generalization of \cite[Theorem~1]{srivastava2024varyingtopologies} for fully-connected networks. In the fully-connected network, one can obtain tight bounds on the propagation speed of a single packet through the network. In our setting, Lemma~\ref{lemma: 1} guarantees bounds on the version age of a vertex. Nevertheless, the same proof technique can be applied.

In our running example, Theorem~\ref{thm: 1} states that if we have two sequences of graphs, one the sequence of fully-connected networks, and the other the sequence of ring networks, the long-term average version age of the varying-topologies network with $n$ nodes has logarithmic scaling if the holding times in both states of the CTMC have mean $O(\log{n})$.

\begin{Proof}
    We immediately obtain the lower bound by replacing $G_{2,n}$ with $G_{1,n}$ and noting this is a gossip network with time-invariant edge rates and long-term version age $\Theta(f_1(n))$.

    We can obtain an upper bound by replacing $G_{2,n}$ by a network with no connections, even with the source. It is easy to see that the disconnected network has the worst long-term average version age of any $n$-vertex network. It now suffices to show that in this modified varying-topologies network, the long-term average version age is $O(f_1(n))$. We proceed by controlling the version age of a single node in this modified process. Without loss of generality, we take this node to be the node labeled 1 throughout the graph sequence. Suppose the version age of node 1 scales as $O(g(n))$ in the single topology network with $G_{1,n}$. We aim to show that it also scales as $O(g(n) + h(n))$ in the varying-topologies network with the disconnected topology.

    Take $n$ and $t_0$ large enough so Lemma~\ref{lemma: 1} holds. For brevity let $G_k = G_{k,n}$ for $k \in \{1,2\}$. Let $\beta > 0$, and let $[s_1, f_1), [s_2,f_2), \ldots,[s_k, f_k)$ be the intervals of time that the CTMC is in state $G_1$, such that $\sum_{i=1}^k (f_i - s_i) = \beta g(n)$ and $s_1 > t_0$. Then, by the stationarity of the CTMC, Lemma~\ref{lemma: 2} and Lemma~\ref{lemma: 3} guarantee the existence of some $\alpha$ depending only on $\beta$ so the event $\mathcal{A} = \{f_k \leq \alpha \log n, N_0[0,f_k] \leq \alpha \log n \}$ holds with high probability.

    If we consider a static gossip network on $G_1$ for total time $T = \sum_{i=1}^k (f_i - s_i)$, then by the observations above, stationarity gives that the distribution of arrivals on any edge is the same in $G_1$ for the interval $[0,T)$ as for the varying-topologies network conditioned on the intervals $[s_i, f_i)$. In the complementary intervals when the CTMC is in the disconnected topology, the age scales as $O(h(n))$. Then, by Lemma~\ref{lemma: 1} and our assumptions on $n$ and $t_0$, the version age of node 1 in the varying-topology network is $O(g(n) + h(n))$. Averaging over all nodes and using the fact that $h(n) = O(f_1(n))$ completes the proof.
\end{Proof}

\section{Typical Sets and Varying-Topologies}\label{sec: typical sets}
\begin{figure}
    \centering
    \includegraphics[width=0.9\linewidth]{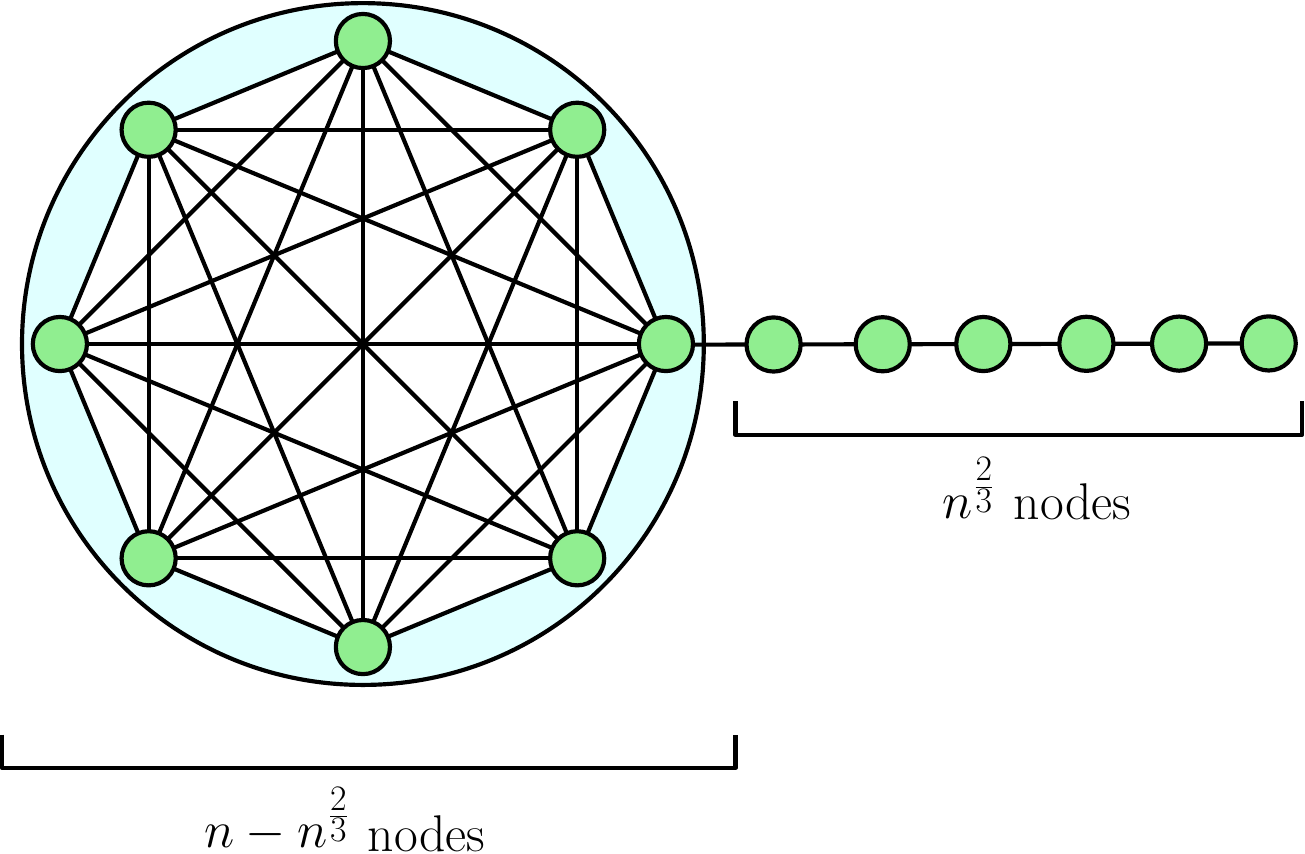}
    \caption{A network formed by combining a fully-connected network with $n-n^{\frac{2}{3}}$ nodes to a line with $n^{\frac{2}{3}}$ nodes.  This network poses challenges to the average-case analysis with slow topology switching.}
    \label{fig: special network topology}
\end{figure}

If the underlying topologies are not vertex-transitive (i.e., vertices may have different neighborhoods) then there can exist a large variation in version age of different nodes. Consider the network illustrated in Fig.~\ref{fig: special network topology}. It can be shown that the long-term average version age for this network is $\Theta(n^{\frac{1}{3}})$. However, nodes in the fully-connected sub-network have $\Theta(\log n)$ expected version age, while nodes in the linear sub-network have $\Theta(n^{\frac{2}{3}})$ expected version age. Thus, if $h(n) = \Theta(n^{\frac{1}{3}})$ and the second topology is fully-connected, the nodes in the linear network never reach the version age they have in the single-topology network. This has unpredictable effects on the long-term average version age of the varying-topologies network. Therefore, the average age analysis from Section~\ref{sec: fast switching} cannot apply to general gossip networks when $h(n) = \Omega(f_1(n))$. This discussion motivates the following definition.

\begin{definition}
    Let $\{G_{n}\}_{n=1}^\infty$ be a sequence of networks with $|\mathcal{N}(G_{k})| = k$. Suppose $G_{n}$ has a long-term average version age $\Theta(f(n))$. Then, the typical set is defined as the set of nodes in $G_{n}$ with long-term average version age $O(f(n))$. Next, suppose $\{G_{1,n}\}_{n=1}^\infty$ and $\{G_{2,n}\}_{n=1}^\infty$ are two sequences of networks with $|\mathcal{N}(G_{1,k})| = |\mathcal{N}(G_{2,k})| = k$. If $G_{1,n}$ and $G_{2,n}$ are two topologies in the varying-topologies network, then the typical set of the varying-topologies network is the intersection of their respective typical sets.
\end{definition}

In our running example, all nodes in the fully-connected topology and the ring topology are a part of the typical set of nodes. In Fig.~\ref{fig: special network topology}, the nodes in the fully-connected part of the network and a few nodes in the linear part of the network are a part of the typical set of nodes. Now, we will show that there are at most $o(n)$ vertices outside the typical set.

\begin{lemma}\label{lemma: 5}
    Let $\{G_{n}\}_{n=1}^\infty$ be a sequence of networks with $|\mathcal{N}(G_{k})| = k$. Suppose $G_{n}$ has a long-term average version age $\Theta(f(n))$. The typical set of $G_{n}$ contains $\Omega(n)$ nodes and its compliment contains $o(n)$ nodes.
\end{lemma}

\begin{Proof}
    Each node which is not in the typical set has $\omega(f(n))$ long-term average version age. Suppose the average version age of the set containing all these nodes is $\Theta(g(n))$ where $g(n) = \omega(f(n))$, and suppose this set has $kn$ nodes, where $0 < k < 1$ is a constant. Similarly, let the average version age of nodes in the typical set be $\Theta(h(n))$, where $h(n) = O(f(n))$. Then, we write the long-term average version age of the entire network in terms of $h(n)$ and $g(n)$.
    \begin{align}
        v^{G_n} =& \frac{h(n)(n-kn)+g(n)kn}{n}\\
        =& h(n)(1-k)+kg(n)\\
        =& \Theta(g(n)).
    \end{align}
    This is in contradiction to the fact that the long-term average version age of the network is $\Theta(f(n))$. Hence, the compliment set has $o(n)$ nodes, and the result follows.
\end{Proof}

The following result is a simple extension of Lemma~\ref{lemma: 5}.

\begin{corollary}
    Suppose we have a varying-topologies network with two topologies. Let $\{G_{1,n}\}_{n=1}^\infty$ and $\{G_{2,n}\}_{n=1}^\infty$ be two sequences of networks with $|\mathcal{N}(G_{1,k})| = |\mathcal{N}(G_{2,k})| = k$. Then, the typical set of nodes in the varying-topologies network with $G_{1,n}$ and $G_{2,n}$ as the two topologies contains $\Omega(n)$ nodes and the remaining nodes are $o(n)$ in number.
\end{corollary}
 
Next, we state and prove the following theorem, which characterizes the behavior of the version age of the typical set of nodes in a varying-topologies network when the holding times are $O(f_1(n))$ or $\Omega(n\log{n})$. We note that for $h(n)$ between $\Omega(f_1(n))$ and $o(n\log{n})$, the long-term average version age is $O(f_2(n))$, which is a universally applicable upper bound.

\begin{theorem}\label{thm: 2}
    Suppose we have a varying-topologies network with two topologies. Let $\{G_{1,n}\}_{n=1}^\infty$ and $\{G_{2,n}\}_{n=1}^\infty$ be two sequences of networks with $|\mathcal{N}(G_{1,k}| = |\mathcal{N}(G_{2,k}| = k$. Suppose $G_{1,n}$ has a long-term average version age $\Theta(f_1(n))$ and $G_{2,n}$ has a long-term average version age $\Theta(f_2(n))$. Further, assume $f_1(n) = o(f_2(n))$ and $q_{12}, q_{21} = \Theta(1/h(n))$. Then, we have the following:
    \begin{enumerate}
        \item If $h(n) = O(f_1(n))$, then the long-term average age scaling for the typical set in the varying-topologies network is $\Theta(f_1(n))$.
        \item If $h(n) = \Omega(n\log{n})$, then the long-term average age scaling for the typical set in the varying-topologies network is $\Theta(f_2(n))$.
    \end{enumerate}
\end{theorem}

In our running example, Theorem~\ref{thm: 2} states that if the holding times have mean smaller than $O(\log{n})$, then the long-term average version age of the typical set of nodes in the varying-topologies network scales as $\Theta(\log{n})$. Further, if the holding times are of order $n\log n$ or larger, then the long-term version age scaling of the typical set of nodes in the varying-topologies network is $\Theta(\sqrt{n})$.

\begin{Proof}
        Item 1) is exactly the case considered in Theorem~\ref{thm: 1}.       
        For item 2), we immediately obtain the upper bound by replacing $G_{1,n}$ with $G_{2,n}$.
        A lower bound can be obtained by lower bounding the version age of nodes during their time spent in $G_{1,n}$ with $0$. As shown in \cite{maranzatto2024agegossipconnectiveproperties}, if a network remains in a fixed topology for a sufficiently long time, the distribution of the version age at any node converges to the distribution of the first-passage time from the source to that node. This is because after a sufficient amount of time, w.h.p. every node in the network receives some update from the source. We now show that $h(n)$ time is long enough for the convergence to happen. Specifically, we obtain an $o(1)$ upper bound for the probability that the source does not send an update to a node within $\Theta(n)$ time. Without loss of generality, let $B_i$ be the random variable associated with the first time that the source sends an update to node $i$ along edge $(n_0, i)$. Let $B = \sup_{i \in \mathcal{N}}B_i$ be the first time every node receives an update from the source. It is known that $\mathbb{E}[B] = nH^{(1)}(n)$ and $Var[B] = n^2H^{(2)}(n)$ where $H^{(k)}(n) = \sum_{j=1}^n \frac{1}{j^k}$. Then, we have the following,
        \begin{align}
            \mathbb{P}\big(|B-\mathbb{E}[B]| > n{\log{n}}\big) \leq \frac{\text{Var}[X]}{(n{\log{n}})^2}
            = \frac{1}{\log^2{n}}
            = o(1).
        \end{align}
        This leads us to conclude that every node receives an update directly from the source in $\Theta(n\log{n})$ time w.h.p. Thus, the distribution of the version age of each node in the typical set of $G_{2,n}$ is the same as the first passage time after $B$. Since $B =  O(h(n))$, the expectation of the version age during any interval spent in $G_{2,n}$ is $O(f_2(n))$ for each node in the typical set, and thus it is also $O(f_2(n))$ w.h.p. with the proof being along similar lines as the proof of Lemma~\ref{lemma: 1}. Thus, the average version age whenever we are in $G_{2,n}$ is $\Theta(f_2(n))$. Similarly, the average version age whenever we are in $G_{1,n}$ is $\Theta(f_1(n))$. From Lemma~\ref{lemma: 2}, we know that we spend a constant fraction of time in each topology w.h.p. Thus, not adding the version age of a node while in $G_{1,n}$ multiplies the long-term average version age of a node by some constant $0 < c < 1$. This gives the lower bound of $\Theta(f_2(n))$.
\end{Proof}

\appendix

\begin{ProofLemma1}
    Choose $t > t_0$ to be chosen later. The version age metric for node 1 at time $t$, $X_1(t)$, can take values only in $\mathbb{N} \cup \{0\}$. Thus, the probability distribution is a probability mass function. Suppose the distribution assigns a cumulative non-zero mass to values $g(n)$ such that $f(n) = o(g(n))$, i.e., $\sum_{i = \omega(f(n))} P(i) > 0$, where $P(i)$ is the probability of the version age of the node being $i$. If this mass is $O(\frac{f(n)}{g(n)})$, then it does not affect the average version age, and our result holds since the mass goes to $0$ when $n$ becomes large. However, if the mass is $\omega(\frac{f(n)}{g(n)})$, then the expected value is $\mathbb{E}[X_1(t)] = \omega(f(n))$. Then, we want to show the contradiction that the long-term average version age of node $1$ cannot be $\Theta(f(n))$. It was shown that the distribution of version age of a node at a particular time instant is the same as the minimum of $t$ and the first passage time $T(1,s)$ between the source and the node. Since we are considering long-term average version age, choose $t_0 = t_0(n)$ so that for all $t > t_0$, $\mathbb{P}(\min{(T(1,s),t)} = T(1,s))=1$. With this choice the distribution is the same for all time instants after $t$. Thus, the long-term average version age $\lim_{t \rightarrow \infty} \mathbb{E}[X_1(t)]$ is also $\omega(f(n))$. This is a contradiction. Thus, there cannot be any non-zero mass larger than $O(\frac{f(n)}{g(n)})$ at $g(n)$. Hence, the version age at any time is $O(f(n))$ w.h.p. when $n$ becomes large.
\end{ProofLemma1}

\begin{ProofLemma2}
    Without loss of generality, let $k = 1$. For some $\beta \geq 0$, let $T$ be the first time that the CTMC has spent a total of $\beta h(n)$ time in state $1$. Then, $T$ is a stopping time. Furthermore, by definition of $T$, we are in state $1$ of the CTMC at time $T$. We may start in state $1$ or state $2$ of the CTMC. If we start in state $1$, we will leave the state $N_{1}(\beta h(n))$ times where $N_{1}$ is the counting process associated with the Poisson process when in state $1$ of the CTMC. If we start in state $2$, we will reach state $1$ and then return to the state $N_{1}(\beta h(n))$ times. We note that the time to return to state $1$ is an inter-arrival time for a renewal process whose arrival times are the instances of return to state $1$. These inter-arrival times, $\{I_i\}_{i=1}^{N_{1}(\beta h(n))}$, follow a two-parameter hypoexponential distribution.
    
    Suppose we spend a total of $T_{out}$ time in state $2$ up to time $T$, then $\text{Var}[T] = \text{Var}[T_{out}]$. Further, let the time of reaching state $1$ for the first time be $T_1$, with $T_1 = 0$ if we start in state $1$. Then, $T_{out} = T_1 + \sum_{i=1}^{N_{1}(\beta h(n))}I_i$. We note that $N_{1}(\beta h(n))+1$  is a stopping time for the Poisson process associated with state $1$ of the CTMC. We also observe that $\mathbb{E}[T_1] \leq \eta \mathbb{E}[I_1]$ for some $\eta \in \mathbb{R}$. Let $\text{Var}[I_i] = \sigma^2$ and $\mathbb{E}[I_i] = \mu$. Then, we have
    \begin{align}
        \mathbb{E}[T_{out}] =& \mathbb{E}[T_1] + \mathbb{E}\bigg[\sum_{i=1}^{N_{1}(\beta h(n))}I_i\bigg]\\
        \leq& (\eta - 1)\mu + \mathbb{E}\bigg[\sum_{i=1}^{N_{1}(\beta h(n))+1}I_i\bigg]\\
        =& (\eta - 1)\mu + \mathbb{E}[N_{1}(\beta h(n))+1]\mu\\
        =& (\eta - 1)\mu + (q_{1} \beta h(n) + 1)\mu\\
        \leq& \rho h(n),
    \end{align}
    for some $\rho \in \mathbb{R}$. The expectation is also lower bounded by $h(n)$ if we remove $T_1$ from the equation and write the expectation. Thus, we let $\mathbb{E}[T_{out}] = \nu h(n)$. Due to the strong Markov property and independence of these inter-arrival times, we can further say, $\text{Var}[T_{out}] = \text{Var}[T_1] + \sum_{i=1}^{N_{1}(\beta h(n))}\text{Var}[I_i]$. Note that $\text{Var}[T_1] \leq \chi\text{Var}[I_0]$, where $\chi \in \mathbb{R}$ and $I_0$ has the same distribution as $I_1$. This is due to the positive recurrence of the CTMC. Thus, we can write $\text{Var}[T_{out}] \leq (\chi^2-1)\text{Var}[I_0] + \sum_{i=1}^{N_{1}(\beta h(n))+1}\text{Var}[I_i]$.

    We know that the variance is finite. Now, we can find the upper bound for $\text{Var}[T]$,
    \begin{align}
        \!\!\!\!\text{Var}[T] =& \text{Var}[T_{out}]\\
        \leq& (\chi^2-1)\text{Var}[I_0] + \sum_{i=1}^{N_{1}(\beta h(n))+1}\text{Var}[I_i]\\
        =& (\chi^2-1)C + \sigma^2 \mathbb{E}[N_{1}(\beta h(n))+1] \notag\\
        &+ \mu \text{Var}[N_{1}(\beta h(n))+1]\\
        =& (\chi^2-1)C + \sigma^2 (q_{1} (\beta h(n))+1) + \mu q_{1} (\beta h(n))\\
        \leq& \xi h(n),
    \end{align}
    for some $\xi \in \mathbb{R}$. Then, using Chebyshev's inequality, we write
    \begin{align}
        \mathbb{P}\big[|T - \mathbb{E}(T)| > h(n)\big] \leq& \frac{\text{Var}[T]}{(h(n))^2}
        \leq \frac{\xi}{h(n)}
        = o(1).
    \end{align}
    Hence, $\beta h(n) + \rho h(n) - h(n) \leq T \leq \beta h(n) + \rho h(n) + h(n)$ with high probability.
\end{ProofLemma2}

\bibliographystyle{unsrt}
\bibliography{refs}

\end{document}